\documentstyle[twocolumn,prl,aps,epsf]{revtex}
\begin{document}
\draft
%\preprint{}

\title{Dynamics of the Peierls-active phonon modes in CuGeO$_3$} 

\author{Claudius Gros and Ralph Werner} 
\address{   Institut f\"ur
 Physik, Universit\"at Dortmund, 44221 Dortmund, Germany.}

\date{\today}

\maketitle

\begin{abstract}
We reconsider the Cross and Fischer approach to
spin-Peierls transitions.
We show that a soft phonon
occurs only if $\ \Omega_0<2.2\,T_{SP}$. For CuGeO$_3$ this
condition is not fulfilled and the calculated temperature
dependence of the Peierls-active phonon modes is in 
excellent agreement with experiment. A central peak
of a width $\sim0.2\,{\rm meV}$ is predicted at $T_{SP}$. 
Good agreement is found between theory and experiment
for the pretransitional Peierls-fluctuations.
Finally, we consider the problem of 
quantum criticality in CuGeO$_3$.
\end{abstract}


%%%%%%%%%%%%%%%%%%%%%%%%%%%%%%%%%%%%%%%%%%%%%%%%%%%%%%%%%%%%%%%%%%%%%%%%
%%%%%%%%%%%%%%%%%%%%%%%%%%%%%%%%%%%%%%%%%%%%%%%%%%%%%%%%%%%%%%%%%%%%%%%%
\centerline{\hfill}
\paragraph*{Introduction}
Structural phase transitions come essentially in two varieties, 
those with a soft phonon mode and those without phonon softening and a
central peak \cite{Bruce}. Typically one associates them to
displacive and to order-disorder transitions respectively, even
though there is no strict formal distinction between
displacive and order-disorder transitions. It came then as a surprise
that the spin-Peierls transition in CuGeO$_3$ \cite{Hase},
which had been shown to be displacive \cite{Hirota}, shows no
phonon softening \cite{Hirota,Lorenzo}. Even worse, the Peierls-active
phonon modes harden by about $5\%-6\%$ with decreasing temperature
\cite{Braden97}. It has been generally assumed, up to now,
that this behaviour is inconsistent with the RPA-approach
by Cross and Fischer (CF) to the spin-Peierls transition 
\cite{CF}.
Here we show, that the CF-theory is actually fully consistent
with the experimental results for CuGeO$_3$. 
We show, that soft phonons occur within RPA only if
the bare phonon frequency $\Omega_0$ satisfies
$\Omega_0<2.2\,T_{SP}$. For larger phonon frequencies
the phonon does not soften and a central peak develops
at the spin-Peierls transition temperature, $T_{SP}$.
We then test the applicability of RPA to CuGeO$_3$ by
calculating the pretransitional Peierls-fluctuations.
We find good agreement with experiment. 
Finally, we note that a key ingredient of the 
CF-approach, quantum criticality, can be tested for
in CuGeO$_3$.

\paragraph*{RPA approach}
The retarded phonon Green's function $D_q(\omega)$
is given by \cite{Mahan}
\begin{equation}
D_q(\omega) = { 2 \Omega_0(q) \over
  \omega^2 - \Omega_0^2(q) - 2\Omega_0(q)P_q(\omega) },
\label{D_q}
\end{equation}
were $\Omega_0(q)$ is the frequency of the 
bare phonon with momentum $q$.
In RPA one approximates the phonon self-energy
$P_q(\omega)$ by $g_q^2\chi_q(\omega)$, where
$\chi_q(\omega)$ is the dynamical energy-energy 
correlation function and where $g_q$ is
the electron-phonon coupling constant, given by
\begin{equation}
|g_{q}|^2={\lambda^2\,\hbar\over M\Omega_0(q)}(1-\cos(qc)),
\label{spin-phonon}
\end{equation}
where we have used
\begin{equation}
\sum_{n} 
       \lambda(u_{n+1}-u_{n})
     {\bf S}_n\cdot{\bf S}_{n+1},\qquad
\label{model}
\end{equation}
for the spin-phonon coupling within a 
linear-chain model.
Here ${\bf S}_n$ are the spin operators at site $n$,
$u_{n}$ the displacement operators
for the normal coordinates of the Peierls-active
phonon-mode \cite{Braden97,Braden96}
and $M$ is the effective mass of the normal mode. 

%%%%%%%%%%%%%%%%%%%%%%%%%%%%%%%%%%%%%%%%%%%%%%%%%%%%%%%%%%%%%%
%
   \begin{figure}[bth]
   \epsfxsize=0.48\textwidth
   \centerline{\epsffile{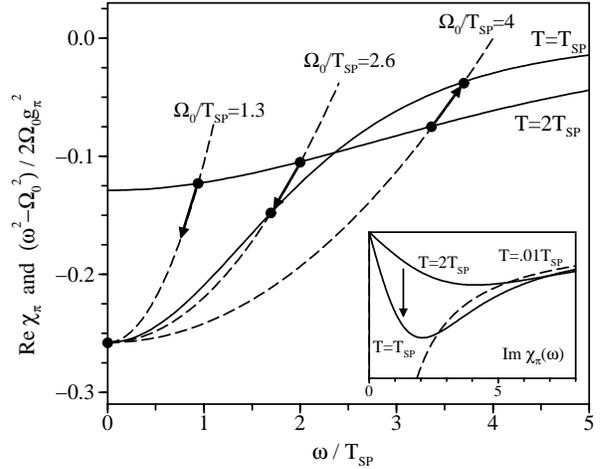}}
   \centerline{\parbox{0.48\textwidth}{\caption{\label{Fig1}
   Plots of Re$\chi_\pi(\omega)$ (solid lines) for
   $T=T_{SP}$ and $T=2T_{SP}$, as a function of $\omega/T_{SP}$.
   The temperature independent LHS of Eq.\ (\protect\ref{poles}),
   $(\omega^2-\Omega_0^2)/(2\Omega_0g_\pi^2)$, 
   is plotted for $\Omega_0/T_{SP}=1.3,\ 2.6,\ 4$
   (dashed lines), with
   $g_\pi$ given by Eq.\ (\protect\ref{T_SP}). The filled
   circles denote the position of the phonon frequencies,
   $\omega_\pi$. The arrows indicate the shift of $\omega_\pi$
   with decreasing temperature.
   Inset: Im$\chi_\pi(\omega)$.
   }}}
   \end{figure}
%
%%%%%%%%%%%%%%%%%%%%%%%%%%%%%%%%%%%%%%%%%%%%%%%%%%%%%%%%%%%%%%%%%

At the spin-Peierls transition a spontaneous dimerization occurs
below $T_{SP}$, at $q=\pi/c$. In the following we set the
lattice constant $c$ to unity in the theory formulas.
Cross and Fischer observed, that the correct functional 
from for $\chi_q(\omega)$ (in the limit $\omega\rightarrow0$)
can be obtained from bosonization \cite{CF},
\begin{equation}
T\,\chi_q(\omega) = -2d\,I_1\left({\omega-\Delta\over2\pi T}\right)
                       \,I_1\left({\omega+\Delta\over2\pi T}\right)
\label{chi}
\end{equation}
where $d\approx0.37$ is a constant depending weakly on
the momentum cut-off, $\Delta=v_s|q-\pi|$ is the
lower edge of the two-spinon continuum ($v_s$ is the
renormalized spin-wave velocity), and
\[
I_1(k)={1\over2\pi}\int_0^\infty{\rm e}^{ikx}(\sinh(x))^{-1/2}.
\]
$T\chi_{q=\pi}(\omega)$ is scale-invariant and a function
of $\omega/(2\pi T)$ only (independent of the
spin-spin coupling $J$). This behaviour is
characteristic of quantum critical systems
\cite{quantum_critical}.
For any temperature $T>0$ we can expand 
$T\chi_\pi(\omega)$ in $\omega/(2\pi T)$ as
\begin{equation} 
T\,\chi_{\pi}(\omega) = -\chi_0 
-i\chi_1 \left({\omega\over2\pi T}\right)
+\chi_2 \left({\omega\over2\pi T}\right)^2
+\dots,
\label{expansion}
\end{equation}
with $\chi_0\approx 0.26$,
$\chi_1\approx 0.81$ and 
$\chi_2\approx 2.2$. The position of the
poles $\omega_\pi$ of $D_\pi(\omega)$ are then determined by
the roots of
\begin{equation} 
{\omega^2-\Omega_0^2\over 2\Omega_0g_\pi^2}
= Re\chi_\pi(\omega)\approx
-{\chi_0 \over T}
+{\chi_2 \over T} \left({\omega\over2\pi T}\right)^2,
\label{poles}
\end{equation}
where $\Omega_0\equiv\Omega_0(\pi)$.
Typical plots of the left and right hand side
of (\ref{poles}) are presented in Fig.\ \ref{Fig1}.

%%%%%%%%%%%%%%%%%%%%%%%%%%%%%%%%%%%%%%%%%%%%%%%%%%%%%%%%%%%%%%
%
   \begin{figure}[bth]
   \epsfxsize=0.48\textwidth
   \centerline{\epsffile{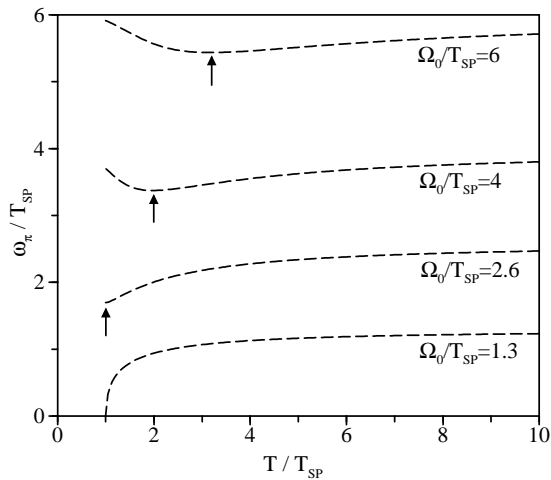}}
   \centerline{\parbox{0.48\textwidth}{\caption{\label{Fig2}
    The temperature dependence of the phonon frequencies
    $\omega_\pi$ for various values of $\Omega_0/T_{SP}$.
    $\Omega_0$ is the bare phonon frequency.
    The arrows indicate the respective minimal phonon
    frequency.
   }}}
   \end{figure}
%
%%%%%%%%%%%%%%%%%%%%%%%%%%%%%%%%%%%%%%%%%%%%%%%%%%%%%%%%%%%%%%%%%

A spontaneous lattice dimerisation, i.e.\ a macroscopic
occupation of the Peierls-active phonon mode, occurs
at $T_{SP}$ when (\ref{poles}) has a solution for $\omega=0$. 
This determines the transition temperature as \cite{CF}
\begin{equation} 
T_{SP} = {2g_\pi^2\over\Omega_0}\chi_0.
\label{T_SP}
\end{equation}
Remarkably, Eq.\ (\ref{T_SP}) is independent of $J$,
due to the scale-invariance of $T\chi_\pi(\omega)$.
We compare the prefactor of the terms $\sim\omega^2$
of the rhs and lhs of Eq.\ (\ref{poles}) 
and find that for
\begin{equation} 
1/(2g_\pi^2\Omega_0) > \chi_2/(4\pi^2T_{SP}^3),
\label{soft_0}
\end{equation}
Eq.\ (\ref{poles}) has a single solution for
$T=T_{SP}$ and inspection of the temperature
dependence of this solution for $T>T_{SP}$
(compare Fig.\ (\ref{Fig1})) shows that this
root continuously connects to the $T=\infty$ solution,
$\ \lim_{T\to\infty}\omega_\pi=\Omega_0(\pi)$.
In the parameter regime defined by Eq.\ (\ref{soft_0}) 
the phonon softens completely. We can use 
Eq.\ (\ref{T_SP}) to eliminate $g_\pi$ from
Eq.\ (\ref{soft_0}). We obtain
\begin{equation} 
T_{SP} > {\Omega_0\over2\pi}\sqrt{\chi_2\over\chi_0}
       \approx 0.46\,\Omega_0,
\qquad\quad \Omega_0<2.2\,T_{SP},
\label{soft_1}
\end{equation}
for the soft-phonon regime.
For $\Omega_0>2.2\,T_{SP}$ the 
Peierls-active phonon does not
soften completely and may even become harder with decreasing
temperature, as illustrated in Fig.\ \ref{Fig2}.
Near $T=T_{SP}$ an additional central peak
shows up, leading to the phase transition.
For CuGeO$_3$ there are two Peierls-active phonon modes
with energies \cite{Braden97} 
$\omega_1=151\,{\rm K}$ and $\omega_2=317\,{\rm K}$ 
respectively. Since (see below) 
$\Omega_\gamma\approx\omega_\gamma$ ($\gamma=1,2$)
and $T_{SP}=14.1\,{\rm K}$ we find that CuGeO$_3$ 
is in the central-peak regime.

\paragraph*{Application to CuGeO$_3$}
In order to compare more in detail with the experimental
results for CuGeO$_3$ we have generalized Eq.\ (\ref{D_q}) 
for the case of two phonon frequencies. Denoting by
$D_{1/2}(\omega)$ the retarded Green's functions of
the first and second phonon with bare frequencies
$\Omega_1$ and $\Omega_2$ respectively, and by $g_1$ and $g_2$
the respective spin-phonon coupling constants, we obtain
\[
D_1(\omega) = D_1^{(0)}(\omega)
            + {\left(D_1^{(0)}(\omega)\right)^2g_1^2\chi_\pi(\omega)\over
1-\left(g_1^2D_1^{(0)}(\omega) + g_2^2D_2^{(0)}(\omega)
  \right)\chi_\pi(\omega)
              }
\]
and an equivalent equation for $D_2(\omega)$. Here
$D_{1/2}^{(0)}(\omega)=2\Omega_{1/2}/(\omega^2-\Omega_{1/2}^2)$
\cite{Mahan}. An analysis similar to the one-phonon case can be 
performed for $D_1(\omega)+D_2(\omega)$. One finds 
\begin{equation} 
T_{SP} = \left({2g_1^2\over\Omega_1}
              +{2g_2^2\over\Omega_2}\right)\chi_0
\label{T_12}
\end{equation}
for the transition temperature and
\begin{equation} 
T_{SP} > \sqrt{\chi_2\over\chi_0}
         {\Omega_1\Omega_2\over2\pi}
         \sqrt{g_1^2\Omega_2+g_2^2\Omega_1\over
               g_1^2\Omega_2^3+g_2^2\Omega_1^3}
\label{soft_2}
\end{equation}
for the soft-phonon regime.

In order to determine $g_1$ and $g_2$ for
CuGeO$_3$ we note that
the lower/upper phonon mode contribute to
the structural distortion below $T_{SP}$ with
weighting factors 2 and 3 respectively \cite{Braden97}.
This leads to
\begin{equation} 
{g_1^2/\Omega_1\over g_2^2/\Omega_2} = {2\over3},\qquad
{g_1^2\over g_2^2} = {2\,\Omega_1\over3\,\Omega_2}
              \approx{2\,\omega_1\over3\,\omega_2}
\approx {1\over3}.
\label{weighting}
\end{equation}
Eq.\ (\ref{weighting}) and Eq.\ (\ref{T_12})
determine the spin-phonon couplings $g_1,\ g_2$.
For $T_{SP}=14.1\,{\rm K}$ we find
$\Omega_1=3.15\, {\rm THz}$ and $\Omega_2=6.61\,{\rm THz}$ for
the bare phonon frequencies and
$g_1=0.86\,{\rm THz}$ ($g_2=\sqrt{3}g_1$) for the
spin-phonon coupling.

\begin{figure}[bth]
   \epsfxsize=0.48\textwidth
   \centerline{\epsffile{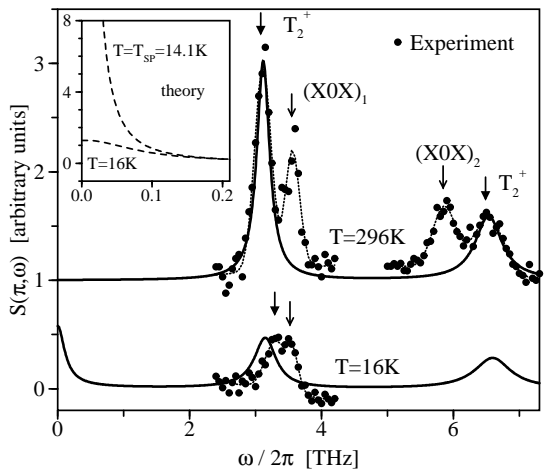}}
   \centerline{\parbox{0.48\textwidth}{\caption{\label{Fig3}
   Theoretical (thick solid lines) and experimental 
   (solid circles, \protect\cite{Braden97}) results for the
   dynamical structure factor. The T$_2^+$ phonons are 
   Peierls active and included in the theory, (XOX)$_1$
   and (XOX)$_2$ are other nearby phonons.
   The data for $T=296\,{\rm K}$ has been shifted.
   Inset: Blowup of the central peak at $T=T_{SP}=14.1\,{\rm K}$
   and $T=16\,{\rm K}$ (not broadened).
   }}}
   \end{figure}

In Fig.\ \ref{Fig3} we have plotted the results for
the dynamical structure factor,
\begin{equation} 
S(\pi,\omega) = -{1\over\pi}{
  Im\left[ D_1(\omega+i\delta) + D_2(\omega+i\delta)\right]
\over 1-\exp(-\beta\omega)},
\label{dnamical}
\end{equation}
where we have used the
experimental resolution function [THz]
$\delta \approx 0.023 + 0.028\omega/(2\pi)$ \cite{note_Braden}.
The intensity of the experimental spectra \cite{Braden97},
also shown in Fig.\ \ref{Fig3}, have been scaled;
the (constant) background has been adjusted
\cite{note_exp}.
The overall agreement between experiment and theory
is satisfactory, although the hardening of the
lower phonon mode is somewhat more pronounced in
the experiment (6\% vs.\ 1\%). No experimental
data for the upper mode were available for $T=16\,{\rm K}$.
In the inset a blowup of the central peak
is given. It should be possible to resolve the
predicted central peak, which has a width of
$\approx0.05\,{\rm THz}=0.2\,{\rm meV}$, by neutron
scattering, testing thereby the theory.

%%%%%%%%%%%%%%%%%%%%%%%%%%%%%%%%%%%%%%%%%%%%%%%%%%%%%%%%%%%%%%%%%
%%%%%%%%%%%%%%%%%%%%%%%%%%%%%%%%%%%%%%%%%%%%%%%%%%%%%%%%%%%%%%
%
%
%%%%%%%%%%%%%%%%%%%%%%%%%%%%%%%%%%%%%%%%%%%%%%%%%%%%%%%%%%%%%%%%%
%%%%%%%%%%%%%%%%%%%%%%%%%%%%%%%%%%%%%%%%%%%%%%%%%%%%%%%%%%%%%%

The theory presented here is based on the RPA approximation
and we have shown it to yield reasonably good
agreement with experiment, explaining the absence of
a Peierls-active soft-phonon mode in CuGeO$_3$.
From a theoretical point of view, one might
question the applicability of RPA in the
central-peak regime. A definite theoretical 
resolution to this problem is not known at
present, but one may note that the
standard phenomenological
theory for the central peak occurring in structural
phase transitions has RPA-form \cite{Bruce}.
For the case of CuGeO$_3$ we will test
the applicability of RPA by comparing 
the prediction of RPA (with no fit parameter) for
the pretransitional Peierls-fluctuations, given by
the inverse lattice correlation length $1/\xi$, 
with the experimental results.

\begin{figure}[bth]
   \epsfxsize=0.48\textwidth
   \centerline{\epsffile{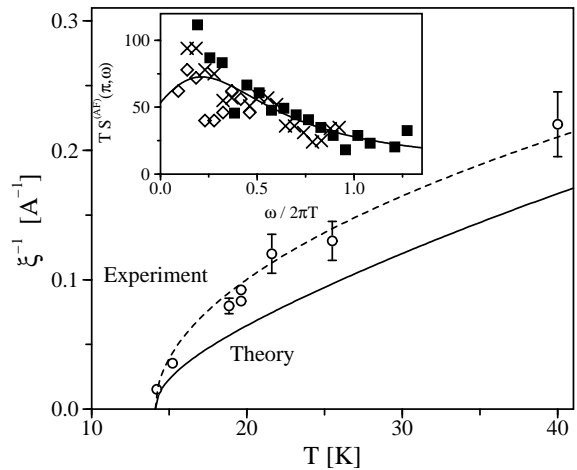}}
   \centerline{\parbox{0.48\textwidth}{\caption{\label{Fig4}
   RPA and experimental results 
   \protect\cite{Pouget94} for the inverse
   lattice correlation length, $1/\xi$.
   The theory does not contain any free parameter.
   Inset: $T\,S^{(AF)}(\pi,\omega)$, 
   as a function of $\omega/(2\pi T)$,
   as predicted by bosonization (solid line,
   Eq.\ (\protect\ref{chi})), and the neutron-scattering
   results \protect\cite{Hirota}, for 
   $T=14.5\,{\rm K}$ (filled squares),
   $T=20\,{\rm K}$ (crosses) and
   $T=50\,{\rm K}$ (diamonds). 
   }}}
   \end{figure}

The lattice correlation length is determined by the
long-distance falloff,
\begin{equation} 
\lim_{z\to\infty}\int {dq\over2\pi} 
{\rm e}^{iqz}
Re D_q(0) 
\ \sim\
{\rm e}^{i \pi z/c} {\rm e}^{-z/\xi},
\label{xi}
\end{equation}
where $c=2.94\,{\rm \AA}$ is the c-axis 
lattice constant of CuGeO$_3$ and 
$D_q(\omega)=\sum_\gamma D_{q,\gamma}(\omega)$.
We have calculated $1/\xi$ from Eq.\ (\ref{xi}),
using 
$v_s=(\pi/2)J(1-1.12\alpha)$ \cite{Fledderjohann}
(which enters Eq.\ (\ref{chi})),
$J=156\,{\rm K}$ for the exchange integral and
$\alpha=0.24$ for the frustration parameter
\cite{Castilla,Muthukumar,gap}.
The results for $1/\xi$ 
are presented in Fig.\ \ref{Fig4}, 
together with results for CuGeO$_3$ obtained
by diffusive X-ray scattering \cite{Pouget94},
which are consistent with neutron-scattering
data and the absence of a soft phonon \cite{Hirota}.

%%%%%%%%%%%%%%%%%%%%%%%%%%%%%%%%%%%%%%%%%%%%%%%%%%%%%%%%%%%%%%%%%
%%%%%%%%%%%%%%%%%%%%%%%%%%%%%%%%%%%%%%%%%%%%%%%%%%%%%%%%%%%%%%
%
%
%%%%%%%%%%%%%%%%%%%%%%%%%%%%%%%%%%%%%%%%%%%%%%%%%%%%%%%%%%%%%%%%%
%%%%%%%%%%%%%%%%%%%%%%%%%%%%%%%%%%%%%%%%%%%%%%%%%%%%%%%%%%%%%%

Both experiment and theory show
mean-field behaviour, $1/\xi\sim\sqrt{T-T_{SP}}$.
The RPA-result agrees well with experiments.
Above $T=19\,{\rm K}$ the lattice fluctuations have 
one-dimensional character \cite{Pouget94} and the
residual difference between theory and experiment
may be due to corrections to RPA.

It is interesting to note, that Eq.\ (\ref{chi}) 
for $\chi_q(\omega)$ is (for
spin-rotational invariant Heisenberg
chains) identical with the bosonization result
for the magnetic dynamical structure factor,
$S^{(AF)}(q,\omega)$ \cite{Schulz86}. 
Quantum criticality implies 
$T\,S^{(AF)}(\pi,\omega)$ to be an 
universal function of $\omega/(2\pi T)$, at least
for small $\omega$.
In the inset of Fig.\ \ref{Fig4} we have plotted
the bosonization result for $T\,S^{(AF)}(\pi,\omega)$,
together with the rescaled neutron-scattering results
\cite{Hirota} 
We observe that the experimental
data approximately obey the scaling, though there is
substantial scattering of the data for small
$\omega/(2\pi T)$, possibly influenced by
non-critical contributions from the Peierls-fluctuations
or by a crossover of the character of
the magnetic excitations from 1D to
2D near the Peierls transition
\cite{Muthukumar}. It is also interesting to note,
that the prediction for  $T\,S^{(AF)}(\pi,\omega)$
is independent of $J$ and that the data for other
1D Heisenberg antiferromagnets with very different
values of the coupling $J$, like KCuF$_3$
\cite{Tennant}, should fall onto the same universal 
curve presented in the inset of Fig.\ \ref{Fig4}. 
An experimental verification of quantum criticality
for $S^{(AF)}(\pi,\omega)$ would imply
also scale-invariance for
$\chi_\pi(\omega)$, since both coincide
within bosonization \cite{CF,Schulz86}.

\paragraph*{Generalization}
Until now we assumed
spin-rotational invarianiance.
Next we will show, that a central-peak
regime occurs also in spin-Peierls 
transitions lacking spin-rotational
invariance. An example is the
spin-Peierls transition in a system
of phonons coupled to an array of chains
with Ising spins. This model was solved
exactly by Pytte \cite{Pytte}. It contains
a parameter regime, where the
transition is displacive and phonons do
not become soft. In the oppositve limit,
when the spin-chains are xy-like, the 
transition corresponds via the
Jordan-Wigner transformation to the
standard Peierls transition \cite{Rice73}.
Again one can show \cite{neu}, that
soft-phonons occur in RPA, e.g.\ for
$T_{SP}=J/10$, only for $\Omega_0<0.8J$.
For $\Omega_0>0.8J$ the Peierls-active
phonon does not become soft and a
central peak arises at $T_{SP}$.

\paragraph*{Discussion}
In this letter we have shown, that the RPA approach
to the spin-Peierls transition includes both a
soft-phonon and a central-peak regime. This
result is at first sight counterintuitive,
as continuous lattice distortions below $T_{SP}$
are generally associated with a softening of the
lattice above $T_{SP}$. 

The eigenstates of the spin-phonon system
evolve adiabatically as a function of the
spin-phonon coupling strength in the soft-phonon
regime.
In the central-peak regime a new magnetophonon
appears at low frequencies and condenses at
$T_{SP}$, leading to the structural transition and
the formation of spin-singlets.
This new collective excitation
is a superposition
of a phonon with two magnons in a singlet state.
The magnetophonon couples to the phonon propagator
and therefore shows up as a low-energy resonance
in $D_q(\omega)$, the central peak. The other resonance in
$D_q(\omega)$, at $\omega_\pi$, has the limit
$\ \lim_{g_\pi\to0}\omega_\pi=\Omega_0(\pi)$.
Therefore, one usually regards $\omega_\pi$ 
to be the ``true'' phonon frequency. In terms
of the eigenstates of the coupled spin-phonon system
such a distinction does not make sense. 
In the central-peak regime the spectral weight 
of $D_q(\omega)$ is divided in between the 
``phonon-resonance'' at $\omega_\pi$
and the soft magnetophonon.

\paragraph*{Conclusions}
The absence of a soft
Peierls-active phonon mode in CuGeO$_3$ has
been considered as a challenge to theory. It has been
argued \cite{Uhrig} that the Cross and
Fischer theory is essentially incomplete,
i.e.\ not applicable to CuGeO$_3$. Here
we point out, that the absence of soft
phonons does actually find a natural
explanation within the CF-approach. 
The calculated temperature-dependence of the
phonon modes and that of the pretransitional
Peierls-fluctuations are in excellent
agreement with experiment. A central
peak of width $0.2\,{\rm meV}$
is predicted to appear at $T_{SP}$.
Finally we have pointed out, that a
key ingredient of the theory, the
quantum criticallity of $\chi_\pi(\omega)$,
can be tested, albeit indirectly, 
with neutron scattering
through a test of the scale invariance
of the magnetic dynamical structure factor, 
$S^{(AF)}(\pi,\omega)$.

We acknowledge discussions with
M. Braden, A. Kluemper, U. L\"ow, V.N. Muthukumar and
W. Weber. We are especially grateful to M. Braden
for sending us the data files.
The support of the German science
foundation is acknowledged.

%%%%%%%%%%%%%%%%%%%%%%%%%%%
%%%%%%%%%%%%%%%%%%%%%%%%%%%

%%%%%%%%%%%%%%%%%%%%%%%%%%%%%%%%%%%%%%%%%%%%%%%%%%%%%%%%%%%%%%%%%
%%%%%%%%%%%%%%%%%%%%%%%%%%%%%%%%%%%%%%%%%%%%%%%%%%%%%%%%%%%%%%%%%
%


\begin{thebibliography}{99}
\bibitem{note_exp} A rough comparsion 
                   \protect\cite{note_Braden}
                   indicates, that the
                   experimental intensity ratio of
                   the lower T$_2^+$ mode
                   at T=295K and T=16K is about 3:1,
                   in agreement with theory, see
                   Fig.\ \protect\ref{Fig3}.
\end{thebibliography}
\end{document}